\documentclass[twocolumn,preprintnumbers,amsmath,amssymb,showpacs,prl,apsrev4-1]{revtex4-1}
\usepackage{graphicx}
\usepackage{dcolumn}
\usepackage{bm}
\usepackage{SIunits}
\usepackage{verbatim}
\usepackage{placeins}

\begin{document}
\title{Orbital-driven two-dome superconducting phases in iron-based superconductors}
\author{Da-Yong Liu$^{1}$,
        Feng Lu$^{2}$,
        Wei-Hua Wang$^{2}$,
        Hai-Qing Lin$^{3}$,
    and Liang-Jian Zou$^{1,4}$}
\altaffiliation{Corresponding author} \email{dyliu@theory.issp.ac.cn; zou@theory.issp.ac.cn}
\affiliation{
      $^1$ Key Laboratory of Materials Physics, Institute of Solid State
      Physics, Chinese Academy of Sciences, P. O. Box 1129, Hefei, Anhui
      230031, China \\
      $^2$ Department of Electronics and Tianjin Key Laboratory of Photo-Electronic Thin Film Device and Technology, Nankai University, Tianjin 300071, China \\
      $^3$ Beijing Computational Science Research Center, Beijing 100193, China \\
      $^4$ Department of Physics, University of Science and Technology of China, Hefei, 230026, China
}

\date{\today}

\begin{abstract}
Recent several experiments revealed that novel bipartite magnetic/superconducting phases widely exist in iron pnictides and chalcogenides. Nevertheless, the origin of the two-dome superconducting phases in iron-based compounds still remains unclear. Here we theoretically investigated the electronic structures, magnetic and superconducting properties of three representative iron-based systems, {\it i.e.} LaFeAsO$_{1-x}$H$_{x}$, LaFeAs$_{1-x}$P$_{x}$O and KFe$_{2}$As$_{2}$. We found that in addition to the degenerate in-plane anisotropic $xz$/$yz$ orbitals, the quasi-degenerate in-plane isotropic orbitals drive these systems entering into the second parent phase. Moreover, the second superconducting phase is contributed by the isotropic orbitals rather than the anisotropic ones in the first superconducting phase, indicating an orbital-selective pairing state. These results imply an orbital-driven mechanism and shed light on the understanding of the two-dome magnetic/superconducting phases in iron-based compounds.
\end{abstract}


\vskip 300 pt

\maketitle


Iron-based superconductors display rich phase diagrams with spin-density wave (SDW) \cite{Nature453-899,EPL83-27006}, orbital order \cite{PRL103-267001,PRB84-064435}, nematicity \cite{science329-824,nphys11-959,PRB92-085109} and superconductivity \cite{JACS130-3296,Nature453-761,PRL101-206404}. These competing or coexisting phases arise from the interplay among charge, spin, orbital and lattice degrees of freedom of iron-based compounds. In the early stage it had already realized that the multi-orbital characters of band structures and orbital ordering in many different families of iron pnictides and chalcogenides \cite{nmat10-932,RMP85-849,PRL100-237003,PRL102-247001,NJP11-025021,PRL102-126401,PRL104-216405,PRL110-146402,RPB92-184512}, and the orbital selective Mott physics was also suggested in some iron-based compounds \cite{PRL102-126401,PRL110-146402}, however, our understand to the roles of orbital degree of freedom seem to be no more than that. The key roles of orbital degree of freedom on superconducting (SC) pairing interactions, pairing symmetry, and phase diagrams in these different iron-based superconductors seem to be far from understood or revealed.

The parent compound LaFeAsO possesses a stripe antiferromagnetic (SAFM) ground state associated with a weak $xz$ orbital order in the low-temperature phase, and exhibits an orthorhombic to tetragonal structural phase transition upon doping or increasing temperature. Once substituting O with F, such an SAFM is gradually suppressed and an SC dome emerges in LaFeAsO$_{1-x}$F$_{x}$ \cite{JACS130-3296}. While for hydrogen doping, recent experiments demonstrated that the phase diagram shows two SC domes accompanied with bipartite magnetic parent phases in LaFeAsO$_{1-x}$H$_{x}$, as seen the AF1-SC1-SC2-AF2 phase diagram in Refs. \cite{ncomms3-943,nphys10-300}. The neutron diffraction experiment showed that the second magnetic parent phase lies around $x$=0.5 in LaFeAsO$_{1-x}$H$_{x}$. This novel parent phase also shows SAFM order except that Fe spin has a large magnetic moment of $\sim$1.21 $\mu_{B}$, in comparison with $\sim$0.63 $\mu_{B}$ in LaFeAsO. Moreover, it possesses an unusual orthorhombic lattice distortion accompanied by the instability of As atoms which are away from the center of the Fe square lattice in the low-temperature phase. These unusual phenomena demonstrate that the orbital physics in LaFeAsO$_{1-x}$H$_{x}$ could be different from LaFeAsO. One naturally arises a question whether the two SC domes in LaFeAsO$_{1-x}$H$_{x}$ is particular? Detailed analysis to recent experimental data shows that some other iron-based compounds also have two distinct AFM/SC phases, demonstrating that such two SC domes may be a general phenomenon in various iron-based SC families.

More recently, several experiments revealed that LaFeAs$_{1-x}$P$_{x}$O exists a different AF1-SC1-AF2-SC2 phase diagram \cite{JPSJ83-023707,JPSJ83-083702,PRB90-064504}. Also in K$_{1-x}$Fe$_{2-y}$Se$_{2}$, a second SC phase re-emerges under high pressure \cite{nature483-67}. In KFe$_{2}$As$_{2}$, a second SC phase is also found accompanied with a collapsed tetragonal phase under high pressure \cite{PRB91-060508}. In addition, two SC domes are also observed in the K-doped FeSe thin films grown on SiC substrate \cite{PRL116-157001}. And in the intercalated FeSe \cite{Sci.Rep.5-9477}, the high pressure also could lead to two SC domes, similar to K$_{1-x}$Fe$_{2-y}$Se$_{2}$. On the other hand, the pressure can induce a new AFM parent phase in pure FeSe \cite{PRB93-094505}. The more recent nuclear magnetic resonance (NMR) experiment showed that there also exist two SC domes and a second orthorhombic phase in LaFeAsO$_{1-x}$F$_{x}$ \cite{CPL32-107401}. More and more bipartite AFM/SC phases have been found in various iron-based compounds, however, the nature of the second AFM/SC phase still remains unclear.

In this Letter, we aim to clarify the origin of the bipartite AFM/SC phases in LaFeAsO$_{1-x}$H$_{x}$, LaFeAs$_{1-x}$P$_{x}$O, and KFe$_{2}$As$_{2}$ compounds. For this purpose, we first study the electronic, magnetic and SC properties of LaFeAsO$_{1-x}$H$_{x}$ using the {\it first-principles} method, projected Wannier functions, the random phase approximation (RPA) and mean-field approximation based on five-orbital Hubbard models. We find that quasi-degenerate $xy$/$xz$/$yz$ orbitals drive the emergence of the second magnetic parent phase, and the in-plane isotropic $xy$-orbital dominates an orbital-selective SC pairing state in LaFeAsO$_{1-x}$H$_{x}$. While in LaFeAs$_{1-x}$P$_{x}$O, one of the quasi-degenerate orbitlas, {\it i.e. } in-plane isotropic $3z^{2}-r^{2}$ orbital, is responsible for the second magnetic parent phase. In KFe$_{2}$As$_{2}$, a quasi-degenerate $xz$/$yz$/$xy$ orbital instability leads to a structural phase transition from a tetragonal phase into a collapsed tetragonal (CT) phase under pressure. Therefore we propose that the quasi-degenerate orbitals play key roles in the emergence of the second AFM phase and the active isotropic orbital controls the second SC phase, leading to the two SC domes in the phase diagrams of these iron pnictides and chalcogenides.

\begin{figure}[htbp]
\hspace*{-2mm}
\centering
\includegraphics[trim = 0mm 0mm 0mm 0mm, clip=true, width=1.0 \columnwidth]{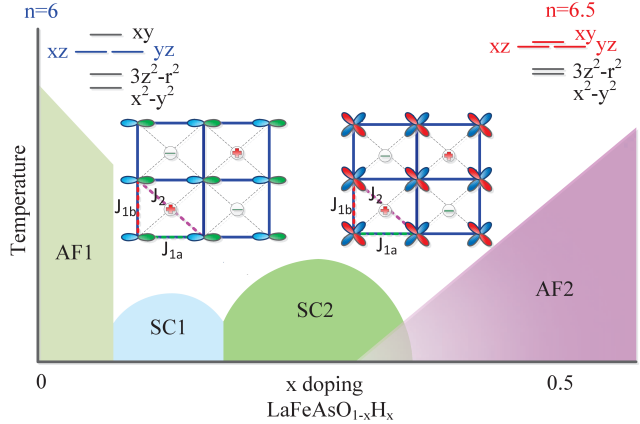}
\caption{(Color online) Schematic phase diagram of LaFeAsO$_{1-x}$H$_{x}$. Different orbital configurations are displayed in the first and second AFM/SC phases.}
\label{pd1}
\end{figure}

\begin{figure}[htbp]\centering
\includegraphics[trim = 0mm 0mm 0mm 0mm, clip=true, width=1.0 \columnwidth]{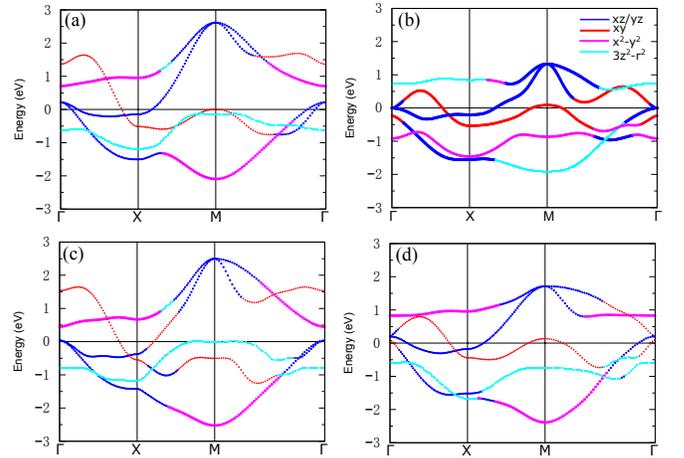}
\caption{(Color online) Orbital-resolved band structures of the five-orbital tight-binding model for (a) LaFeAsO, (b) LaFeAsO$_{0.5}$H$_{0.5}$, (c) LaFeAs$_{0.5}$P$_{0.5}$O and (d) LaFePO.}
\label{bs}
\end{figure}

%
Fig.~\ref{pd1} schematically displays the phase diagram of LaFeAsO$_{1-x}$H$_{x}$ according to the recent experiments \cite{ncomms3-943,nphys10-300,PRB91-064509} and our numerical calculations and analysis. We find that the degenerate $xz$/$yz$ orbitals dominate the AF1 phase at $x$$\sim$0 and the SC1 phase, while the quasi-degenerate $xy$/$xz$/$yz$ orbitals dominate the AF2 phase at $x$$\sim$0.5 and the SC2 phase, respectively.
In unveiling the different orbital physics in LaFeAsO$_{1-x}$H$_{x}$, we first study the electronic structures of two parent phases LaFeAsO$_{1-x}$H$_{x}$ at $x$=0 and 0.5, respectively. With the help of the projected Wannier function technique, the five-orbital tight-binding model $H_{0}$ is constructed. The orbital-resolved band structures are displayed in Fig.~\ref{bs} (a) and (b). It is clearly showed that besides the $xz$/$yz$ orbitals, the $xy$ orbital also contributes considerable weight near Fermi level ($E_{F}$) at $x$=0.5, different from that at $x$=0. For comparison, the crystal field splittings, {\it i.e.} on-site energies of Fe-3$d$ orbitals in two compounds can be seen in Supplemental Material \cite{suppl} and Ref. \cite{PRL101-087004}.

The most obvious multi-orbital feature of LaFeAsO$_{1-x}$H$_{x}$ at $x$=0.5 is that the three $t_{2g}$ orbitals, $xz$/$yz$ and $xy$, have nearly identical on-site energies, thus are almost degenerate. We call these three orbitals {\it quasi-degeneracy}, since the bandwidths of these three orbitals in LaFeAsO$_{0.5}$H$_{0.5}$ are clearly distinguished with $W_{xz/yz}>W_{xy}$.
The anion-height instability, which was recently suggested to be responsible for the AF2 phase \cite{PRL112-187001}, is in fact a natural consequence of the orbital quasi-degeneracy; according to the Jahn-Teller effect, these quasi-degenerate orbitals are unstable, resulting in the lattice distortion at low temperatures. On the other hand, due to different orbital components and weights in the two parent phases with $x$=0 and 0.5, a single rigid-band tight-binding model could not describe the entire $T$-$x$ phase diagram. Thus to depict the low-energy physics in LaFeAsO$_{1-x}$H$_{x}$, we have to adopt two sets of different tight-binding parameters at $x$=0 for AF1 and SC1 phases, and at $x$=0.5 for AF2 and SC2 phases, respectively.

In order to explore the role of Fermi surface nesting and the possible magnetic instability in LaFeAsO$_{1-x}$H$_{x}$ at $x$=0.5, we study the dynamical spin susceptibility within RPA. The bare spin susceptibility is given by the formula \cite{NJP11-025016,RPP74-124508,RMP84-1383}
\begin{eqnarray}
  \chi_{0}^{l_{1}l_{2}l_{3}l_{4}}(q, i\omega)&=&-\frac{1}{N}\sum_{\substack{\vec{k},\mu\nu}}
  \frac{a_{\nu}^{l_{4}}(\mathbf{k})a_{\nu}^{l_{2},*}(\mathbf{k})a_{\mu}^{l_{1}}(\mathbf{k+q})a_{\mu}^{l_{3},*}(\mathbf{k+q})}
  {\omega+\varepsilon_{\mu}(\vec{k}+\vec{q})-\varepsilon_{\nu}(\vec{k})+i\eta} \nonumber \\
  &&\times[f(\varepsilon_{\mu}(\vec{k}+\vec{q}))-f(\varepsilon_{\nu}(\vec{k}))]
  \label{eq-chi0}
\end{eqnarray}
The RPA susceptibility is expressed as
\begin{eqnarray}
  \chi_{c(s)}^{RPA}(\mathbf{q}, i\omega) &=& \chi_{0}(\mathbf{q},i\omega)[\mathbb{I}\pm\Gamma_{c(s)}\chi_{0}(\mathbf{q},i\omega)]^{-1},
\end{eqnarray}
where $\chi_{0}$ is the bare susceptibility defined in Eq.~\ref{eq-chi0}, and the nonzero components of the matrices $\Gamma_{c}$ and $\Gamma_{s}$ are given as $(\Gamma_{c})_{aa,aa}=U$, $(\Gamma_{c})_{aa,bb}=2U'-J_{H}$, $(\Gamma_{c})_{ab,ab}=-U'+2J_{H}$, $(\Gamma_{c})_{ab,ba}=J_{H}$ and $(\Gamma_{s})_{aa,aa}=U$, $(\Gamma_{s})_{aa,bb}=J_{H}$, $(\Gamma_{s})_{ab,ab}=U'$, $(\Gamma_{s})_{ab,ba}=J_{H}$ with orbitals $a\neq b$.
It is found that the peaks of the dynamical spin susceptibility appear at $Q\sim$($\pi$, $\pi/3$), rather than ($\pi$, 0), implying that the nesting scenario fails in LaFeAsO$_{0.5}$H$_{0.5}$. This is due to the absence of the hole pocket at $\Gamma$ point like K$_{1-x}$Fe$_{2-y}$Se$_{2}$ \cite{PRL106-187001}, in contrast to that in LaFeAsO \cite{suppl,PRB88-041106,PRB93-195148}.

To further understand the AF2 parent phase, we also investigate the magnetic ground state properties of LaFeAsO$_{0.5}$H$_{0.5}$ by the {\it first-principles} electronic structures calculations. We find that a stable magnetic ground state is SAFM, in agreement with the neutron diffraction experiment \cite{nphys10-300}.
Based on the total energy differences among different magnetic configurations and fitting by the $J_{1a}$-$J_{1b}$-$J_{2}$ Heisenberg model, we estimate the spin exchange parameters: $J_{1a}$=7.8 meV/$s^{2}$, $J_{1b}$=4.7 meV/$s^{2}$, and $J_{2}$=31.4 meV/$s^{2}$ in LaFeAsO$_{0.5}$H$_{0.5}$. The next nearest-nearest-neighbor (N.N.N.) coupling $J_{2}$ is much larger than the N.N. spin couplings $J_{1a}$ and $J_{1b}$, very different from other iron-based compounds. This strong magnetic anisotropy indicates the possible existence of strong orbital order, which we will show in the following.

To explore the interplays among the charge, spin and orbital degrees of freedom including the electronic correlation in AF1 and AF2 parent phases, we consider the five-orbital Hubbard models $H=H_{0}+H_{I}$ for $x$=0 and for $x$=0.5, respectively, to investigate the electronic and magnetic properties. Within the mean-filed approximation, we define the order parameters $n_{\alpha}$ and $m_{\alpha}$ with $\alpha$=$xz$, $yz$, $xy$, $x^{2}-y^{2}$ and 3$z^{2}-r^{2}$ as
\begin{eqnarray}
 n_{\alpha}=\sum_{\substack{k,\sigma}}\langle c_{k,\alpha,\sigma}^{\dag}c_{k,\alpha,\sigma} \rangle, m_{\alpha}=\sum_{\substack{k,\sigma}}\sigma\langle c_{k+\mathbf{Q},\alpha,\sigma}^{\dag}c_{k,\alpha,\sigma} \rangle
\end{eqnarray}
where $\mathbf{Q}$ is magnetic vector and $\sigma$=$\pm$1. Then we can obtain the magnetic moment $m=\sum_{\alpha}m_{\alpha}$ and total particle number $n=\sum_{\alpha}n_{\alpha}$ through the self-consistently calculations.

\begin{figure}[htbp]
\hspace*{-2mm}
\centering
\includegraphics[trim = 0mm 0mm 0mm 0mm, clip=true, width=0.85 \columnwidth]{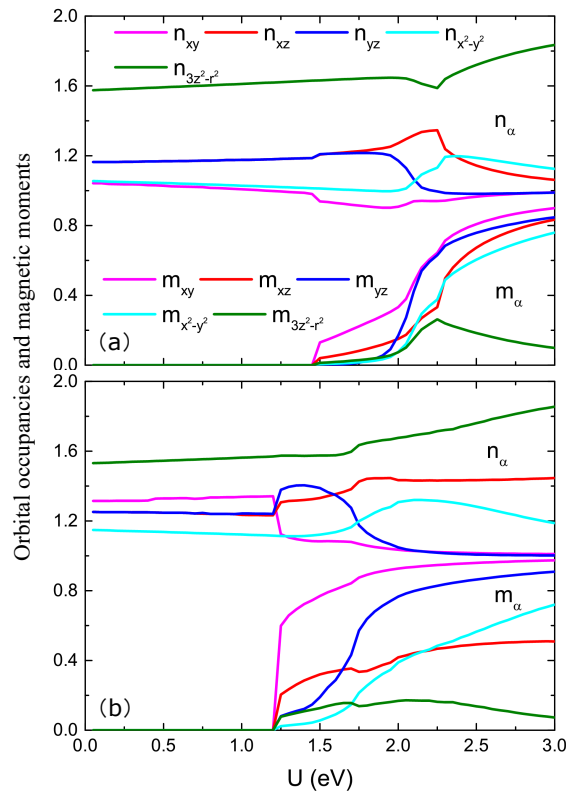}
\caption{(Color online) Orbital occupancies and magnetic moments dependence on Coulomb interaction $U$ within the five-orbital Hubbard model for (a) LaFeAsO and (b) LaFeAsO$_{0.5}$H$_{0.5}$ with parameter $J_{H}$=0.1$U$.}
\label{oomm}
\end{figure}

We determine the magnetic ground state at the mean-field level through considering the following magnetic vectors $\mathbf{Q}$ including (0, 0), ($\pi$, 0), ($\pi$, $\pi$), ($\pi$, $\pi$/3) and (2$\pi$/3, 0). The consequence is that the SAFM phase with $\mathbf{Q}$=($\pi$, 0) is the most stable, in consistent with our preceding LDA data and the experimental results \cite{nphys10-300}. The dependence of the orbital occupancies and magnetic moments on Coulomb interaction $U$ are shown in Fig.~\ref{oomm} (a) for $x$=0 and Fig.~\ref{oomm} (b) for $x$=0.5, respectively. It is obviously found that in LaFeAsO$_{1-x}$H$_{x}$ at $x$=0.5 the orbital polarization occurs in three orbitals when the system enters into the second magnetic parent phase. The original quasi-degenerate three orbitals with almost equal particle number undergo a significant reconstruction, resulting in a slight polarization $n_{o1}=n_{xz}-n_{yz}$ ($n_{xz}<n_{yz}$) but a very large one $n_{o2}=n_{xy}-(n_{xz}+n_{yz})/2$. The large $n_{o2}$ indicates a strong $xy$ ferro-orbital order compared with the weak $xz$-orbital order, as seen the inset of Fig.~\ref{pd1}. The strong $xy$ ferro-orbital order thus leads to a very large N.N. spin exchange parameter $J_{2}$, which explains the strong anisotropy between $J_{1}$ and $J_{2}$ \cite{PRL113-027002}. Therefore, the orbital order scenario involved with three quasi-degenerate orbitals in LaFeAsO$_{1-x}$H$_{x}$ ($x$=0.5) is significantly different from the weak orbital order in LaFeAsO.

From Fig.~\ref{oomm} (b), one notices that in LaFeAsO$_{0.5}$H$_{0.5}$, when $U>U_{c}$$\sim$1.6 eV, $n_{xy}$=$n_{xy\uparrow}+n_{xy\downarrow}$$\sim$1 and $m_{xy}$=$n_{xy\uparrow}-n_{xy\downarrow}$$\sim$1, indicating $n_{xy\uparrow}\sim$1 and $n_{xy\downarrow}\sim$0, {\it i.e.} the spin-up channel of $xy$ orbital is almost fulfilled. This demonstrates that an orbital-selective Mott phase (OSMP) occurs with the insulating $xy$ orbital and the metallic $xz$ and $yz$ orbitals, as seen the sketch of the OSMP \cite{suppl}. The insulating $xy$ orbital exhibits strong ordering with respect to $xz/yz$ orbitals, similar to manganites and other transition-metal oxides. This OSMP accompanied by a strong orbital order contributes to the second AFM parent phase at $x$=0.5.

%
To look into the emergence of the second SC phase SC2, we investigate the dependence of orbital order parameters on doping $x$ within the five-orbital Hubbard model of LaFeAsO$_{1-x}$H$_{x}$ at 0.5 \cite{suppl}. These two orbital ordering parameters show a competing behavior: the $xz$-orbital weight increases while the $xy$-orbital weight decreases when $x$ varies from 0.5 to 0.35, favoring a paramagnetic metallic phase and the emergence of SC2 around $x$$<$0.5 end. Out of expectation, when $x$$>$0.5, the $xy$-orbital weight becomes more large, consistent with the experimental observation \cite{nphys10-300,PRB91-064509}. These results explain why the SC2 phase appears at the left end of AF2 phase in the phase diagram of LaFeAsO$_{1-x}$H$_{x}$.

%
In analyzing the SC pairing properties of the SC2 phase, we calculate the pairing vertex arising from the exchange of spin and orbital fluctuations using RPA. The singlet orbital vertex function $\Gamma_{\alpha\beta\mu\nu}$ is given by
\begin{eqnarray}
  \Gamma_{\alpha\beta\mu\nu}(\mathbf{k},\mathbf{k'},\omega)=[\frac{3}{2}U^{s}\chi_{s}^{RPA}(\mathbf{k-k'}, \omega)U^{s}+\frac{1}{2}U^{s} \nonumber\\
  -\frac{1}{2}U^{c}\chi_{c}^{RPA}(\mathbf{k-k'},\omega)U^{c}+\frac{1}{2}U^{c}]_{\alpha\beta\mu\nu}.
\end{eqnarray}
\begin{figure}[htbp]\centering
\includegraphics[trim = 0mm 0mm 0mm 0mm, clip=true, angle=0, width=1.0 \columnwidth]{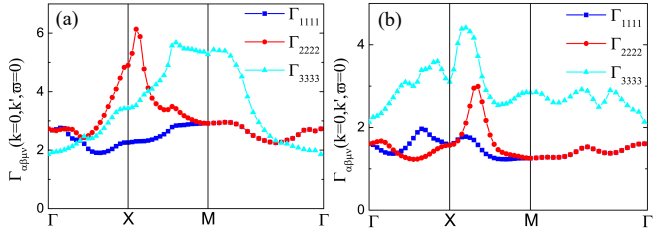}
\caption{(Color online) Orbital pairing vertices along high symmetric directions for (a) LaFeAsO$_{1-x}$H$_{x}$ at $x$=0.125 with $U$=1.4 and $J_{H}$=0.1$U$, and (b) LaFeAsO$_{1-x}$H$_{x}$ at $x$=0.35 with $U$=0.8 and $J_{H}$=0.1$U$. Orbital index 1 refers to $xz$, 2 to $yz$, and 3 to $xy$ orbital, respectively.}
\label{opv}
\end{figure}
The dominant orbital pairing vertices for LaFeAsO$_{1-x}$H$_{x}$ at $x$=0.125 and $x$=0.35 are shown in Fig.~\ref{opv}. It is found that in SC1 phase, the $yz$-orbital dominates the pairing vertex around ($\pi$, 0), while the $xy$-orbital in SC2 phase, implying an orbital-selective pairing state observed by a recent ARPES experiment in iron chalcogenides \cite{ncomms6-7777}. It reveals that the in-plane anisotropic $xz$/$yz$ orbitals in Fe square plane are dominant in SC1 phase, but the in-plane isotropic $xy$ orbital is in SC2 phase. Therefore the orbital fluctuations due to the three quasi-degenerate orbitals also strongly contribute to the SC2 phase.

\begin{figure}[htbp]\centering
\includegraphics[trim = 0mm 0mm 0mm 0mm, clip=true, width=1.0 \columnwidth]{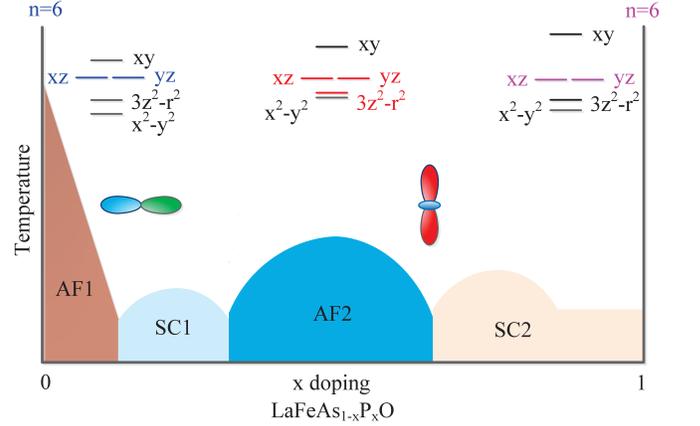}
\caption{(Color online) Schematic orbital configurations in the phase diagram of LaFeAs$_{1-x}$P$_{x}$O.}
\label{pd2}
\end{figure}

%
%
For another two-dome compound LaFeAs$_{1-x}$P$_{x}$O with the isovalent substitution of phosphorus for arsenic, we briefly discuss its origin of two AFM/SC domes. Fig.~\ref{pd2} shows a sketch of the LaFeAs$_{1-x}$P$_{x}$O phase diagram according to the experiments \cite{JPSJ83-023707,JPSJ83-083702,PRB90-064504}. Similar to LaFeAsO$_{1-x}$H$_{x}$, we propose that the degenerate $xz$/$yz$-orbitals are dominant in the AF1 and SC1 phases, while an active in-plane isotropic $3z^{2}-r^{2}$ orbital dominates the AF2 and SC2 phases, which is consistent with the experimental observations \cite{JPSJ83-083702}.
The orbital-resolved band structures are shown in Fig.~\ref{bs} (c) and (d) for LaFeAs$_{1-x}$P$_{x}$O at $x$=0 and 0.5, respectively. From Fig.~\ref{bs} (d) it is clearly found that in the AF2 phase, in addition to the degenerate $xz$/$yz$ orbitals, there also exist quasi-degenerate $3z^{2}-r^{2}$/$x^{2}-y^{2}$ orbitals at $x$$\sim$0.5, which could be seen from the crystal field splittings, {\it i.e.} the on-site energy differences, of Fe-3$d$ orbitals \cite{suppl}).

\begin{figure}[htbp]\centering
\includegraphics[trim = 0mm 0mm 0mm 0mm, clip=true, width=1.0 \columnwidth]{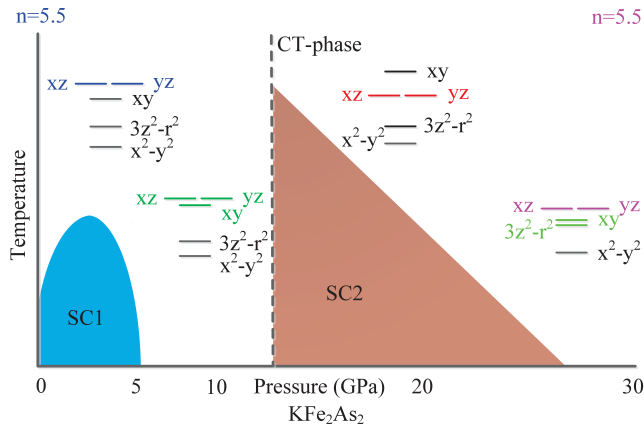}
\caption{(Color online) Schematic orbital configurations in the phase diagram of KFe$_{2}$As$_{2}$ under pressure.}
\label{pd3}
\end{figure}

%
%
Finally, we also perform similar analysis for KFe$_{2}$As$_{2}$ with $n$=5.5 under pressure, and find that the quasi-degenerate character of orbitals occurs when the SC1 or SC2 phase disappears, as seen from the schematic phase diagram in Fig.~\ref{pd3} according to the experimental result in Ref.~\cite{PRB91-060508}. For comparison, the on-site energies of Fe-3$d$ orbitals are given in Supplemental Material \cite{suppl}. The instability of the quasi-degenerate $xz$/$yz$/$xy$ orbitals at about 10 GPa drives a structural phase transition from a tetragonal phase to a CT phase under high pressure up to 15 GPa. The dominant $xy$-orbital character is presented at 10 GPa, but absent at 20 GPa, which is also verified by a recent DFT and dynamical mean-field-theory (DMFT) calculations \cite{PRB91-140503}. In addition, we predict that another quasi-degenerate $xz$/$yz$-$xy$/$3z^{2}-r^{2}$ orbital instability emerges above 30 GPa.

Through the analysis of the above three representative systems, we conclude that the active quasi-degenerate orbitals drive the emergence of the second AFM or SC phase \cite{suppl}. Meanwhile, the orbital physics of the fully degenerate $xz/yz$ orbitals is the dominant scenario in the AF1/SC1 phase \cite{suppl}.
Actually, the doping/substitution (chemical internal pressure) or hydrostatic pressure pushes the in-plane anisotropic $xz/yz$ orbitals with $C$2 symmetry away from $E_{F}$, but it activates the in-plane isotropic $xy$, $x^{2}-y^{2}$ or $3z^{2}-r^{2}$ orbital with $C$4 symmetry in the planar square Fe lattice. Therefore, it is readily understood that the isotropic orbitals emerge around $E_{F}$. The in-plane isotropic orbital would suppress ($\pi$, 0) spin fluctuations, but enhance the ($\pi$, $\pi$) spin fluctuations and orbital fluctuations. Consequently, the orbital fluctuations in the second AF/SC phase play a dominant role, which is supported by a quite recent NMR experiment \cite{arXiv1609.04957}.

%
In summary, we have shown that in addition to conventional magnetic and superconducting phases in LaFeAsO$_{1-x}$H$_{x}$, LaFeAs$_{1-x}$P$_{x}$O and KFe$_{2}$As$_{2}$, the quasi-degenerate orbitals drive the emergence of the second AFM/SC phases, lead to universal two-dome SC phase in iron pnictides. Due to the orbital modulation, there exist two distinct types of SC phases, one is in-plane anisotropic orbital dominant SC1 phase with low $T_{c}$, the other is in-plane isotropic orbital dominant SC2 phase with high $T_{c}$. Actually, our scenario could be generally extended to the similar two-dome SC phases observed experimentally in iron chalcogenides, such as the FeSe-based compounds under pressure, as well as the heavy fermion superconductors. The understanding of novel orbital-selective magnetic/SC state definitely sheds light on the origin of the two-dome phases in iron-based materials. This suggests that, to search for much higher $T_{c}$ iron-based SC materials, the presence of the active isotropic orbital near $E_{F}$ is an important factor.
%

%
This work was supported by the National Sciences Foundation of China under Grant Nos. 11574315, 11274310, 11474287 and 11404172. H. Q. Lin acknowledges financial support from NSAF U1530401. Numerical calculations were performed at the Center for Computational Science of CASHIPS and the Beijing Computational Science Research Center.

\end{document}